\documentclass[reprint,aps,superscriptaddress,nofootinbib,groupedaddress,preprintnumbers]{revtex4-1}

\usepackage{amsmath,amsthm,amsfonts}
\usepackage{graphicx}
\usepackage{physics}
\usepackage{hyperref}
\hypersetup{
	colorlinks   = true, 
	urlcolor     = blue, 
	linkcolor    = blue, 
	citecolor    = red   
}

\graphicspath{{figs/}}

\newcommand{\nc}{\newcommand}

\nc{\calR}{{\cal{R}}}
\nc{\calP}{{\cal{P}}}
\nc{\cN}{ {\cal{N}} }

\nc{\Mpt}{M_{_{\rm Pl}}^2}

\usepackage{tikz}
\usetikzlibrary{arrows,shapes}
\usetikzlibrary{trees}
\usetikzlibrary{matrix,arrows} 				
\usetikzlibrary{positioning}				
\usetikzlibrary{calc,through}				
\usetikzlibrary{decorations.pathreplacing}  
\usepackage{pgffor}							

\usetikzlibrary{decorations.pathmorphing}	
\usetikzlibrary{decorations.markings}
\tikzset{
    vector/.style={decorate, decoration={snake}, draw},
	provector/.style={decorate, decoration={snake,amplitude=2.5pt}, draw},
	antivector/.style={decorate, decoration={snake,amplitude=-2.5pt}, draw},
    fermion/.style={draw=black, postaction={decorate},
        decoration={markings,mark=at position .55 with {\arrow[draw=black]{>}}}},
    fermionbar/.style={draw=black, postaction={decorate},
        decoration={markings,mark=at position .55 with {\arrow[draw=black]{<}}}},
    fermionnoarrow/.style={draw=black},
    gluon/.style={decorate, draw=black,
        decoration={coil,amplitude=4pt, segment length=5pt}},
    scalar/.style={dashed,draw=black, postaction={decorate},
        decoration={markings,mark=at position .55 with {\arrow[draw=black]{>}}}},
    scalarbar/.style={dashed,draw=black, postaction={decorate},
        decoration={markings,mark=at position .55 with {\arrow[draw=black]{<}}}},
    scalarnoarrow/.style={dashed,draw=black},
    electron/.style={draw=black, postaction={decorate},
        decoration={markings,mark=at position .55 with {\arrow[draw=black]{>}}}},
	bigvector/.style={decorate, decoration={snake,amplitude=4pt}, draw},
}

\tikzstyle{block} = [draw, rectangle, 
    minimum height=3em, minimum width=6em]

\begin{document}

\preprint{IFT-UAM/CSIC-24-44}

\title{Criticality and thermodynamic geometry of quantum BTZ black holes}

\author{Seyed Ali Hosseini Mansoori,$^{1}$ Juan F. Pedraza$^{2}$ and Morteza Rafiee$^{1}$}
\email{shosseini@shahroodut.ac.ir\\j.pedraza@csic.es\\m.rafiee@shahroodut.ac.ir}
         
\affiliation{$^{1}$Faculty of Physics, Shahrood University of Technology, P.O. Box 3619995161 Shahrood, Iran}
\affiliation{$^{2}$Instituto de Física Teórica UAM/CSIC, Calle Nicolas Cabrera 13-15, Madrid 28049, Spain}

\begin{abstract}
Within the framework of extended black hole thermodynamics, where the cosmological constant acts as the thermodynamic pressure and its conjugate as the thermodynamic volume, we analyze the phase structure and thermodynamic geometry of the three-dimensional quantum-corrected BTZ (qBTZ) black hole. Our results uncover two-phase transitions in the $T-S$ plane across all pressures except at a critical value. Numerical analysis reveals continuous critical phenomena along the coexistence curve, with critical exponents of 2 and 3 for the heat capacity at constant pressure and the NTG curvature, respectively. Importantly, these values notably deviate from the well-known critical exponents observed in mean-field Van der Waals (VdW) fluids, where the NTG curvature and heat capacity demonstrate discontinuous criticality. To our knowledge, our investigation is the first exploration of critical behavior in black holes incorporating consistent semiclassical backreaction.

\end{abstract}

\maketitle 

\section{Introduction}\label{sec0}

In the past decade, the field of extended black hole thermodynamics, or black hole chemistry, has garnered significant attention \cite{Kastor:2009wy,Dolan:2010ha,Cvetic:2010jb,Kubiznak:2012wp,Kubiznak:2014zwa}. A key advancement in this area is the interpretation of the cosmological constant as a thermodynamic pressure and its conjugate as a thermodynamic volume. This conceptual advancement has led to numerous theoretical developments in understanding the thermodynamics and phase transitions of AdS black holes in the extended phase space, as evidenced by various studies \cite{Altamirano:2013ane,Altamirano:2013uqa,Altamirano:2014tva,Johnson:2014yja,Dolan:2014cja,Wei:2012ui,Cai:2013qga,Sherkatghanad:2014hda,Caceres:2015vsa,Karch:2015rpa,Pedraza:2018eey,Visser:2021eqk,Dutta:2022wbh,Amo:2023bbo}. See \cite{Kubiznak:2016qmn} for a comprehensive review.

In recent years, it has been recognized that the geometry of thermodynamic phase space, characterized by the Ruppeiner metric, provides us with a valuable tool to explore black hole phase transitions and criticality \cite{ruppeiner1979thermodynamics,oshima1999riemann,ruppeiner1995riemannian,aaman2003geometry,mirza2007ruppeiner,Ghosh:2020kba,Dutta:2021whz}.  In this context, a novel formalism of Ruppeiner geometry, dubbed New Thermodynamic Geometry (NTG) \cite{Mansoori:2013pna,HosseiniMansoori:2019jcs,Mansoori:2016jer,Mansoori:2014oia}, has been recently proposed, establishing a one-to-one correspondence between phase transition points and singularities of the corresponding scalar curvature. 

The formalism of NTG not only addresses a notable limitation inherent in the conventional Ruppeiner approach \cite{Sarkar:2006tg,aaman2003geometry} but also reveals additional compelling insights \cite{HosseiniMansoori:2020jrx,Rafiee:2021hyj}. Specifically, a universal behavior has been observed in the NTG curvatures of charged AdS black holes, akin to that of a Van der Waals fluid within the framework of extended black hole thermodynamics. Regardless of spacetime dimension, these black holes consistently exhibit a critical exponent of 2 and a universal critical amplitude of $-1/8$ for their normalized NTG curvatures near the critical point \cite{HosseiniMansoori:2020jrx,Wei:2019uqg,Wei:2019yvs}.

Thus far, criticality results have been confined to the realm of classical black holes in AdS space, corresponding to holographic duals of large-$N$ field theories. However, our primary objective in this study is to initiate an exploration of the phase structure and critical phenomena within holographic theories extending beyond the strict infinite-$N$ limit. In fact, exploring beyond the realm of infinite-$N$ holds promise, as in the classical limit, all extensive quantities scale with the number of degrees of freedom and, when properly organized, extended thermodynamics seemingly represents a straightforward extension of standard thermodynamics \cite{Karch:2015rpa,Visser:2021eqk}. To this end, we shift our focus towards investigating the criticality of the so-called quantum BTZ (qBTZ) black hole \cite{Emparan:2020znc}, a model derived from a specific braneworld construction incorporating consistent semi-classical backreaction. 
 
In the context of braneworld holography \cite{Randall:1999vf,Randall:1999ee,Karch:2000ct,Karch:2000gx,de2001gravity}, quantum black hole solutions on a Randall-Sundrum or Karch-Randall brane arise from classical solutions of the higher dimensional bulk's Einstein equations, with appropriate brane boundary conditions \cite{Emparan:1999wa,Emparan:1999fd,Emparan:2002px,Panella:2024sor}. From the brane perspective, these `quantum' black holes are shown to solve semi-classical gravitational equations, capturing backreaction effects from quantum conformal matter on the brane. Furthermore, braneworld holography offers a robust and natural framework for exploring the extended thermodynamics of black holes \cite{Frassino:2022zaz}. Specifically, variations in brane tension result in a dynamic cosmological constant on the brane, which can be interpreted as pressure within the context of extended thermodynamics.

The remainder of the paper is organized as follows. In Section \ref{sec:background} we give a brief overview of some background material, including elements of braneworld holography and the qBTZ black hole. In section \ref{sec1}, focusing on the qBTZ solution, we derive the heat capacities at constant pressure and volume using the bracket method, alongside presenting the phase structure of the qBTZ on the $T-S$ plane. In Section \ref{sec2}, we employ the NTG geometry to investigate phase transitions and validate their correspondence with thermodynamic curvature singularities. Notably, the phase structure of the qBTZ black hole exhibits characteristics akin to a Van der Waals transition, featuring distinct phases at low and high temperatures. Consequently, in Section \ref{sec3}, we investigate the thermodynamic behavior along the coexistence curve and we determine numerically the critical exponents and amplitudes for thermodynamic quantities such as heat capacities and thermodynamic curvatures, zooming in near the critical point. Finally, in Section \ref{sec4} we provide a summary of our findings and draw conclusions based on our results. 

\section{Basics of braneworld holography and the quantum BTZ black hole\label{sec:background}}

To ensure this article is self-contained, here we include essential background material on braneworld holography and the qBTZ solution. For a more in-depth description of the setup, we refer the reader to \cite{Panella:2024sor}.

Consider standard AdS$_{d+1}/\text{CFT}_{d}$. The bulk manifold is denoted by $\mathcal{M}$ and has a curvature radius $L_{d+1}$. As usual, quantum fluctuations of the CFT induce UV divergences, which can be regularized by adding appropriate counterterms \cite{Kraus:1999di,deHaro:2000vlm,Skenderis:2002wp}, a process known as holographic renormalization. In braneworld holography, a $d$-dimensional brane $\mathcal{B}$ replaces the regulator surface, akin to a Randall-Sundrum or Karch-Randall construction \cite{Randall:1999vf,Randall:1999ee,Karch:2000ct,Karch:2000gx}. This brane serves as a physical cutoff, effectively regulating UV divergences. Moreover, the brane's metric is dynamical; it is located away from the boundary and at such position non-normalizable modes of bulk fields (including the metric) are allowed to fluctuate.

This system is described by the sum of an Einstein–Hilbert action, the Gibbons–Hawking–York boundary term, and a brane contribution:
\begin{equation}\label{BulkTheory}
I_{\text{Bulk}}[\mathcal{M}]+I_{\text{GHY}}[\partial\mathcal{M}]+I_{\text{Brane}}[\mathcal{B}]\,,
\end{equation}
where
\begin{eqnarray}
\!\!I_{\text{Bulk}}&\!\!=\!\!&\frac{1}{16\pi G_{d+1}}\int_{\mathcal{M}} d^{d+1}x\sqrt{-g}\left(\hat{R}-2\Lambda_{d+1}\right)\,,\\
\!\!I_{\text{GHY}}&\!\!=\!\!&\frac{1}{8\pi G_{d+1}}\int_{\partial\mathcal{M}}d^{d}x\sqrt{-h}K\,,\\
\!\!I_{\text{Brane}}&\!\!=\!\!&-\tau \int_{\mathcal{B}}d^{d}x\sqrt{-h}\,.
\end{eqnarray}
Here, $\hat{R}$ and $\Lambda_{d+1} = -d(d-1)/(2L_{d+1}^2)$ represent the bulk's Ricci scalar and cosmological constant, respectively, $K$ is the trace of the extrinsic curvature of $\partial\mathcal{M}$, and $\tau$ denotes the tension of the brane.

Integrating out the bulk between $\partial\mathcal{M}$ up to $\mathcal{B}$ amounts to removing CFT degrees of freedom above the UV cutoff. This leads to the following brane effective action $I$ \cite{Panella:2024sor}
\begin{equation}
I=I_{\text{Bgrav}}[\mathcal{B}]+I_{\text{CFT}}[\mathcal{B}]\,,
\end{equation}
which includes an induced gravity theory,
\begin{eqnarray}
\begin{split}
I_{\text{Bgrav}}&=\frac{1}{16\pi G_{d}}\int_{\mathcal{B}} d^{d}x\sqrt{-h}\biggr[R-2\Lambda_{d}\\
&+\frac{L_{d+1}^{2}}{(d-4)(d-2)}(R^2\text{-terms})+\cdots\biggr]\,,
\end{split}
\end{eqnarray}
coupled to a large-$c$ CFT with a UV cutoff, $I_{\text{CFT}}$, arising from normalizable modes. Here, $L_d$ represents the AdS radius on the brane, and the $R^{2}+...$ terms denote higher curvature contributions that can, in principle, be solved for perturbatively to arbitrary order. The relation between bulk and brane parameters is:
\begin{eqnarray}
G_{d}&=&\frac{d-2}{2L_{d+1}}G_{d+1} \,,\label{eq:effGd}\\
\frac{1}{L_{d}^2}&=&\frac{2}{L_{d+1}^2}\left(1-\frac{4\pi G_{d+1}L_{d+1}}{d-1}\tau\right)\,. \label{eq:effLd}
\end{eqnarray}

The system has three equivalent descriptions, and is thus an example of double holography: (i) the bulk, consisting of Einstein gravity in $(d+1)$-dimensions coupled to a $d$-dimensional brane, (ii) 
the brane, consisting of a higher curvature gravity theory coupled to quantum conformal matter, both in $d$-dimensions, and (iii) the CFT description, consisting of a BCFT in $d$-dimensions coupled to a DCFT in $(d-1)$-dimensions. Importantly,  classical solutions to the bulk theory map to \emph{exact} solutions to the semi-classical equations of motion on the brane,
\begin{equation}
G_{\mu\nu}+\Lambda_d g_{\mu\nu}+\cdots=8\pi G_d \left\langle T_{\mu\nu}\right\rangle_{\text{CFT}}\,,\label{eq:semiEE}
\end{equation}
where the dots represent terms higher in the curvature.

For our purposes, it is important to note that specific bulk black holes map to semi-classical black holes on the brane. A concrete example is the quantum BTZ solution introduced in \cite{Emparan:2020znc}, which is achieved by embedding an AdS$_{3}$ Karch-Randall brane into a bulk AdS$_{4}$ C-metric,
\begin{eqnarray}
\begin{split}
ds_{\text{4D}}^{2}&=\frac{l^{2}}{(l+xr)^{2}}\biggr[-H(r)dt^{2}+\frac{dr^{2}}{H(r)}\\
&\qquad\qquad\qquad+r^{2}\left(\frac{dx^{2}}{G(x)}+G(x)d\phi^{2}\right)\biggr]\,,
\end{split}
\label{eq:Cmet}
\end{eqnarray}
with
\begin{equation} H(r)=\kappa+\frac{r^{2}}{l_{3}^{2}}-\frac{\mu l}{r}\;,\quad G(x)=1-\kappa x^{2}-\mu x^{3}\;.\label{eq:Hfunc}
\end{equation}
Here $\kappa=\pm1,0$ denotes the type of slicing, $l$ is the inverse acceleration of the black hole, and $\mu$ is a positive parameter related to the black hole's mass. Further,  
$l_{3}$ is a length scale related to the curvature scale via 
$L_{4}^{-1}=(\frac{1}{l^{2}}+\frac{1}{l_{3}^{2}})^{1/2}$.

The most advantageous geometric feature of the C-metric (\ref{eq:Cmet}) is that the
$x = 0$ timelike hypersurface is umbilic, for any $\kappa$ and $\mu$. This means that the extrinsic curvature $K_{ij}$ of the hypersurface is proportional to its induced metric $h_{ij}$. This property simplifies the embedding of an AdS$_{3}$ Karch-Randall brane. Specifically, one finds that the Israel junction conditions,
\begin{equation}
\Delta K_{ij}-h_{ij}\Delta K=-8\pi G_4\tau h_{ij}\,,
\end{equation}
are trivially satisfied at $x=0$  if the tension is tuned to
\begin{equation}
\tau =\frac{1}{2\pi G_4 l}\,.\label{eq:branetenqBTZ}
\end{equation}
The metric on the brane  yields the qBTZ solution,
\begin{equation}
\begin{split}
&ds^2_{\text{3D}}=-f(r)dt^2+\frac{dr^2}{f(r)}+r^2d\phi^2\,,\label{eq:qBTZmetric}\\
&f(r)=\frac{r^2}{l_3^2}-8\mathcal{G}_3M-\frac{l F(M)}{r}\,.
\end{split}
\end{equation}
Here, the constant $l_{3}$ can be identified with the AdS$_{3}$ length scale, $\mathcal{G}_3=G_{3}/\sqrt{1+(l/l_{3})^{2}}$ is the `renormalized' Newton's constant (with its bare value being $G_{3}$), $M$ is a constant that we will later identify as the mass of the black hole (\ref{Mfunction}), and $F(M)$ is another constant that can be thought of as a function of the mass, given by
\begin{equation}
\mathcal{F}(M) =8\frac{1-\kappa x_{1}^{2}}{(3-\kappa x_{1}^{2})^{3}}\,,
\end{equation}
where $x_1$ is the smallest  positive root of $G(x)$.

Notably, the qBTZ black hole is guaranteed to be an \emph{exact} solution to the semi-classical Einstein's equations (\ref{eq:semiEE}), to all orders in backreaction. The backreaction parameter is $\nu\equiv l/l_3$, which, at leading order, is approximately $\nu\sim2c G_3/l_3$. When $\nu=0$, (\ref{eq:qBTZmetric}) reduces to the classical BTZ black hole. However, $\nu$ does not need to be perturbatively small. In fact, given that $c$ is large, the corrections to the BTZ solution can be significantly larger than the Planck length $\ell_P\sim G_3$.

\section{Thermodynamics of the quantum BTZ black hole}\label{sec1}
As reviewed in Sec. \ref{sec:background}, the qBTZ black hole arises as an induced black hole within a Karch-Randall AdS$_3$ brane embedded in a slice of AdS$_4$ C-metric \cite{Emparan:2020znc}. The brane's position is characterized by a parameter $l$, inversely related to the brane tension and the black hole's acceleration in the 4-dimensional bulk.  However, in the regime of ‘small acceleration’, there is no acceleration horizon and we have only a black hole horizon in thermal equilibrium with its surrounding. Consequently, the thermodynamics of the qBTZ solution are directly inherited from the thermodynamics in the higher dimensional bulk. 

A direct method to study the thermodynamic behavior of this solution is to compute the on-shell Euclidean action and identify it with the canonical free energy. This was first accomplished in \cite{Kudoh:2004ub} and was recently revisited in \cite{Panella:2024sor}. The net result of the calculation is
\begin{equation}
\!I_{\text{on-shell}}=-\frac{\pi l_3 z\sqrt{1+\nu^2}[1+2\nu z+\nu z^{3}(2+\nu z)]}{G_{3}(2+3\nu z+\nu z^{3})(1+3z^{2}+2\nu z^{3})}\;,
\end{equation}
where the two variables $(z,\nu)$ are defined as:
\begin{equation}
z\equiv \frac{l_{3}}{r_{+} x_{1}}, \qquad \nu\equiv \frac{l}{l_{3}},
\end{equation}  
and both quantities have a range of $[0,\infty)$. Roughly speaking, $z$ controls the temperature of the black hole, while $\nu$ the strength of backreaction due to the CFT$_{3}$ on the brane. Furthermore, $r_{+}$ denotes the location of the event horizon, and $x_{1}$ represents the $x_1$ is the smallest positive root of $G(x)$ appearing in the C-metric.

With the on-shell Euclidean action at hand, we can then compute the mass and entropy for the qBTZ black hole as $M=(\partial I_{\text{on-shell}}/\partial \beta)_{\nu,\l_3}$, $S=\beta(\partial I_{\text{on-shell}}/\partial \beta)_{\nu,l_3}-I_{\text{on-shell}}$, where $\beta$ is the period of the Euclidean time circle, needed to avoid a conical singularity. This yields
\begin{eqnarray}\label{Mfunction}
M(z,\nu)&=&\frac{\sqrt{1+\nu^2}}{2G_{3}} \frac{z^2 (1-\nu z^3)(1+\nu z)}{(1+3 z^2 +2 \nu z^3)^2},\\ \label{Sfunction}
S(z, \nu)&=& \frac{\pi l_{3} \sqrt{1+ \nu ^2}}{G_{3}} \frac{z}{1+3 z^2+ 2 \nu z^3},
\end{eqnarray}
consistent with \cite{Kudoh:2004ub,Johnson:2023dtf,Frassino:2023wpc}.
In the framework of extended thermodynamics of quantum black holes \cite{Frassino:2022zaz}, there are two additional important thermodynamic variables:
the (brane) cosmological constant $\Lambda_{3}=-1/L_3^2$, which is interpreted as the pressure, 
 \begin{equation}\label{Pfunction}
 P\equiv -\frac{\Lambda_{3}}{8 \pi G_3}=\frac{1+\nu^2-\sqrt{1+\nu^2}}{4 \pi G_3 l_{3}^2\nu^2},
 \end{equation}
and the central charge of the CFT backreacting onto the geometry, which is holographically given by
 \begin{equation}\label{centralC}
 c\equiv \frac{L_4^2}{G_4}=\frac{l_3\nu}{2G_3\sqrt{1+\nu^2}}.
 \end{equation}
Notably, from the perspective of the higher dimensional bulk, varying the brane cosmological constant is equivalent to varying the tension $\tau$. This is possible if one simultaneously changes the acceleration of the bulk black hole, as seen from the Israel junction condition (\ref{eq:branetenqBTZ}).
 
All the above quantities satisfy the extended first law
\begin{equation}\label{firstlaw}
dM=T dS+V dP+\mu dc,
\end{equation}
and the semi-classical Smarr relation\footnote{Note that the mass term is absent from the Smarr relation since $G_3M$ has vanishing
scaling dimension in three-dimensions.}
\begin{equation}
0=TS-2PV+\mu c\,,
\end{equation}
 where the temperature $T=\beta^{-1}$, the conjugate volume $V$ and chemical potential $\mu$ satisfy (for more details on the bracket notation, please refer to Appendix \ref{AppA}):
 \begin{eqnarray}\label{Tfunction}
 T&=&\left(\frac{\partial M}{\partial S}\right)_{P,c}=\frac{\{M,P,c\}_{z,v,c}}{\{S,P,c\}_{z,v,c}}\\
 &=&\frac{1}{2 \pi l_{3} } \frac{z(2+3 \nu z+\nu z^3)}{1+3 z^2+2 \nu z^3},\nonumber\\
 V&=&\left(\frac{\partial M}{\partial P}\right)_{S,c}= \frac{\{M,S,c\}_{z,v,c}}{\{P,S,c\}_{z,v,c}}\\
 \nonumber  &=&\frac{-2 \pi l_{3}^2(-2 + \nu^2 + 3 \nu^3 z^3 + \nu z (\nu^2-4)+\nu^4 z^4)}{(1+3 z^2+ 2 \nu z^3)^2},\label{eq:volume}\\
 \mu&=&\left(\frac{\partial M}{\partial c}\right)_{S,P}= \frac{\{M,S,P\}_{z,v,c}}{\{c,S,P\}_{z,v,c}}\\
 \nonumber &=& \frac{-z^2 (1+\nu^2)}{l_{3} \nu^3 (1+z^2 (3+2z \nu))^2}\Big[\nu^2(2+z(3+z^2) \nu)\\
\nonumber  &+&2( \sqrt{1+\nu^2}-1)\Big(\nu^2+z \nu^3+3 z^3 \nu^3+z^4 \nu^4-4 z \nu -2 \Big)\Big].
 \end{eqnarray}
The above quantities are consistent with those obtained in Refs. \cite{Johnson:2023dtf,Frassino:2023wpc}.\footnote{The convention used in the definition of mass in (\ref{Mfunction}) includes a zero-point contribution. Consequently, the volume derived in (\ref{eq:volume}) may become negative for certain parameter ranges. This issue can be fully circumvented by subtracting off the Casimir contribution, thus shifting the definition of $V$ by a constant \cite{Frassino:2024bjg}.} Note that by allowing pressure variations in the first law, the mass function $M$ plays the role of the enthalpy, namely $M=H=E+P V$ where $E$ is the internal energy. The Gibbs free energy, defined by $G=H-TS$, defines the extended phase space.\footnote{In extended thermodynamics the on-shell Euclidean action is identified with the Gibbs free energy (rather than the Helmholtz free energy): $I_{\text{on-shell}}=\beta G$ \cite{Dolan:2010ha}. However, the relations (\ref{Mfunction})-(\ref{Sfunction}) are unchanged because these derivatives are taken at fixed $P$.}

Throughout this paper, we will assume $c$ is a constant, equivalent to working in the `fixed charge ensemble.'
The heat capacity at constant pressure is given by
\begin{eqnarray}\label{CP}
C_{P,c}&=&T\Big(\frac{\partial S}{\partial T}\Big)_{P,c}=T \frac{\{S,P,c\}_{z,\nu,c}}{\{T,P,c\}_{z,\nu,c}}\\
\nonumber &=&\frac{c z \pi (1+\nu^2)  (3 z^2-1+4 \nu z^3)(2+3 \nu z+ \nu z^3)}{\nu (1-\nu z^3)(1+3 z^2+ 2 \nu z^3)(3 z^2-1-3 \nu z + \nu z^3)},
\end{eqnarray}
consistent with the results of \cite{Johnson:2023dtf,Frassino:2023wpc}. Similarly, the heat capacity at constant volume is given by
\begin{eqnarray}\label{CV}
C_{V,c}&=&T\Big(\frac{\partial S}{\partial T}\Big)_{V,c}=T \frac{\{S,V,c\}_{z,\nu,c}}{\{T,V,c\}_{z,\nu,c}}\\
\nonumber &=& \frac{-4 c \pi  z (1+\nu^2) (1+\nu z)^3 (2+\nu z(3+z^2))f(z,\nu)}{(1+z^2(3+2 \nu z)) g(z,\nu)},
\end{eqnarray}
where 
\begin{eqnarray}
f(z,\nu)&=&4 z +\nu (-3+z(-3 z+\nu (3+z^2(3+4 \nu z)))),\\
\nonumber g(z,\nu)&=&16-48z^2+80 \nu z (1-2 z^2)+4 \nu^8 z^8 (z^2-3)\\
\nonumber &+& 8 \nu^7 z^7 (3 z^2-5)
-4 \nu^3 z (3-4 z^2+45 z^4)\\
\nonumber &-& 4 \nu^2 (1-27 z^2 +56 z^4)+ 4 \nu^5 z (3+2 z^2-15 z^4+2 z^6)\\
\nonumber &-&4 \nu^4 (3 z^2-1+17 z^4 +21 z^6)
\\
\nonumber &+& 3 \nu^6 z^2 (3+5 (z^2-3 z^4+3 z^6)).
\end{eqnarray}
Note that the both above heat capacities are linearly dependent on the central charge.

In order to explore the extended phase behavior of the qBTZ black hole, we need to determine the critical point by solving the following pair of conditions
\begin{eqnarray}
\left(\frac{\partial T}{\partial S}\right)_{P}&=&\frac{\{T,P,c\}_{z,\nu,c}}{\{S,P,c\}_{z,\nu,c}}=0\\
\nonumber & \Rightarrow & (\nu z^3 -1)(3 z^2-1+ \nu z (z^2-3))=0,\\
\left(\frac{\partial^2 T}{\partial S^2}\right)_{P}&=&\frac{\{(\partial_{S} T)_{P},P,c\}_{z,\nu,c}}{\{S,P,c\}_{z,\nu,c}}=0\\
\nonumber &\Rightarrow &  (z^2-1) (1+z^2 (3+2 \nu z))^4=0.
\end{eqnarray}
Hence, the critical point is found to be $z=1$ and $\nu=1$, or equivalently,\footnote{Note that qBTZ black holes only exist up to a pressure $P_{\text{max}}=1/16 \pi c^2 G_{3}^3>P_c$ \cite{Frassino:2023wpc}, reached when $\nu \to \infty$.}
\begin{eqnarray}
P_{c}&=&\frac{2-\sqrt{2}}{32 \pi c^2 G_{3}^3}, \hspace{0.2cm} T_{c}=\frac{1}{4 \pi \sqrt{2} c G_{3}}, \hspace{0.2cm} V_{c}=0,\\
\nonumber S_{c}&=&\frac{2 \pi c }{3} , \hspace{0.3cm} \mu_{c}=-\frac{1}{6 \sqrt{2} cG_{3}}.
\end{eqnarray}
It is convenient to define the reduced variables as $\hat{P}=P/P_{c}$, $\hat{T}=T/T_{c}$, $\hat{S}=S/S_{c}$, and $\hat{\mu}=\mu/\mu_{c}$. However, it is not possible to define $\hat{V}$ in this way, since $V_c$ vanishes. Hence, it is more favorable for us to work in the $\hat{T}-\hat{S}$ plane, instead of the more standard $\hat{P}-\hat{V}$ plane, used in studies of extended thermodynamics.

\begin{figure}
\includegraphics[trim={1mm 1mm 0 0},clip,scale=0.6]{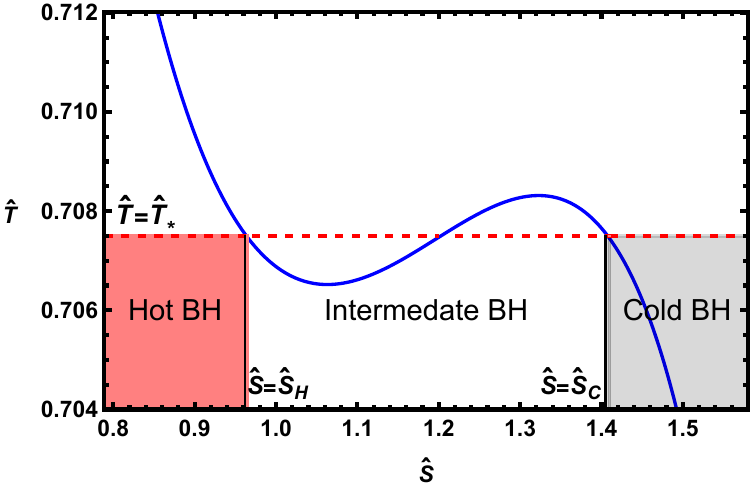}
\caption{A typical $\hat{T}-\hat{S}$ diagram for $\hat{P}<\hat{P}_c$ ($\hat{P}=0.6$). A qualitatively similar diagram can be observed for pressures higher than its critical value, however, when the critical pressure is attained, the Maxwell areas vanish. \label{PT1}}
\end{figure}

As shown in Fig. \ref{PT1}, except for the critical pressure $P=P_{c}$, we observe three distinct phases on the $\hat{T}-\hat{S}$ plane. These include: (a) the cold black hole (CBH), (b) the intermediate black hole (IBH), and (c) the hot black hole (HBH) (we adopt the terminology introduced in \cite{Frassino:2023wpc}, which differs from that of \cite{Emparan:2020znc}). In fact, there exists a first order phase transition between cold and hot black holes. Just like how small-large black hole phase transitions occur in charged and rotating AdS black holes, a similar phenomenon can be observed here. Notice that the critical point is $(\hat T, \hat S)=(1,1)$ in $\hat{T}-\hat{S}$ plane.

\section{Thermodynamic geometry}\label{sec2}

In this section, we employ a novel formalism of thermodynamic geometry, known as NTG \cite{HosseiniMansoori:2019jcs}, to explore the phase behavior of the system. This framework facilitates the identification of a direct mapping between phase transitions and singularities within the thermodynamic scalar curvature. The NTG geometry is characterized by

\begin{equation}\label{Ru1}
dl_{NTG}^2=\frac{1}{T}\left( \eta_i^{ j} \, \frac{\partial^2\Xi}{\partial X^j \partial X^l} \, d X^i d X^l \right),
\end{equation}
where $\eta_i^{ j}={\rm diag} (-1,1,...,1)$, $\Xi$ is a thermodynamic potential and $X^{i}$ are intensive/extensive variables \cite{HosseiniMansoori:2019jcs}. 

As mentioned before, the NTG results extend further our knowledge of phase transition points.  For example, in the fixed central charge ensemble, as one chooses thermodynamic potential by Legendre transformation like $\Xi=H(S,P)=E+P V=M(S,P)$, the  curvature singularity occurs exactly at the same location as the phase transition point of $C_{P,c}$ happens \cite{HosseiniMansoori:2019jcs}. In addition, this result holds true for conjugate potential pairs $(\Xi, \overline{\Xi})$ that satisfy the following relation \cite{HosseiniMansoori:2020yfj},
\begin{equation}
\Xi+\overline{\Xi}=2 E-TS+PV,
\end{equation}
Therefore, one can prove that the singularities of the scalar curvatures $R^{H}$ and $R^{F}$, associated with the conjugate pair $\Xi= M(S,P)=H(S,P)$ and $\bar{\Xi}=F(T,V)=E-TS=M-PV-TS$, respectively, correspond to the divergences of $C_{P,c}$. In addition, their associated metrics are negative of each other, i.e.
\begin{equation}
g^{NTG}_{H}=-J^{T} g^{NTG}_{F} J \hspace{0.5cm} \text{or} \hspace{0.5cm} dl^{2}(H)=-dl^{2}(F),
\end{equation}
where $J$ is the Jacobian matrix defined as 
\begin{equation}
J=\frac{\partial(T,V)}{\partial (S,P)}=\left( \begin{matrix}
\Big(\frac{\partial T}{\partial S}\Big)_{P}& \Big(\frac{\partial T}{\partial P}\Big)_{S}\\
\Big(\frac{\partial V}{\partial S}\Big)_{P} & \Big(\frac{\partial V}{\partial P}\Big)_{S}\\
\end{matrix}\right).
\end{equation}

\begin{figure}[t]
\includegraphics[trim={5mm 1mm 0 0},clip,scale=0.448]{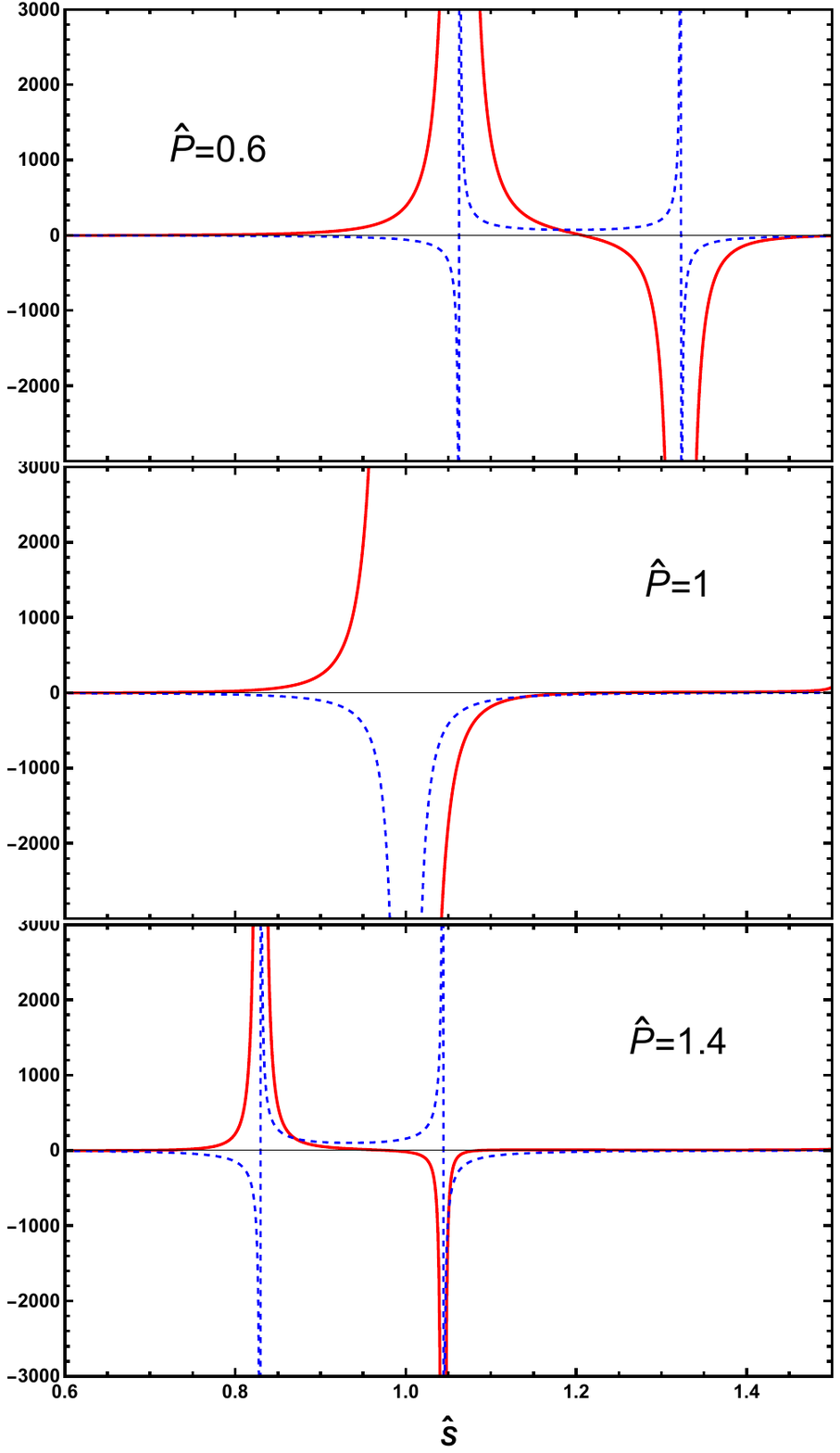}
\caption{The diagram of the specific heat $C_{P,c}$ (blue dashed curve) and  curvature $R^{H}$ (red curve) versus the entropy $\hat{S}$ for $\hat{P}=0.6$, $\hat{P}=1$, and $\hat{P}=1.4$. Note that the same figures can be obtained when one consider $R^{F}$. \label{RHCP1}}
\end{figure}
Equipped with the enthalpy potential $M(S,P)$, we can employ NTG geometry to examine phase transition points of $C_{P,c}$. By substituting  thermodynamic potential $\Xi=H=M=E+PV$  with $X^{i}=(S,P)$ into Eq. (\ref{Ru1}), we arrive at
\begin{eqnarray}\label{met1}
 g^{NTG}_{H}=\frac{1}{T}\text{diag}\left(M_{SS},-M_{PP} \right).
\end{eqnarray}
As all thermodynamic parameters are expressed as a function of $(z,\nu)$, it is convenient to transform the metric elements from the coordinate $X^{i}=(S,P)$ to the preferred coordinate $(z,\nu)$. To do this, we first need to redefine metric elements as follows,
\begin{eqnarray}
M_{SS}&=&\frac{\partial^2 M}{\partial S^2}=\left(\frac{\partial T}{\partial S}\right)_{P}=\frac{\{T,P\}_{z,\nu}}{\{S,P\}_{z,\nu}},\\
M_{PP}&=&\frac{\partial^2 M}{\partial P^2}=\left(\frac{\partial V}{\partial P}\right)_{S}=\frac{\{V,S\}_{z,\nu}}{\{P,S\}_{z,\nu}},
\end{eqnarray}
 then changing from coordinate $(S,P)$ to $(z,v)$ by using the below Jacobian matrix,
 \begin{equation}
 J_{c}=\frac{\partial (S,P)}{\partial (z,\nu)},
 \end{equation}
and finally  convert the metric elements to 
 \begin{equation}\label{met2}
 \hat{g}=J_{c}^{T} g^{NTG}_{M} J_{c}.
 \end{equation}
We have depicted the scalar curvature $R^{H}$ and heat capacity $C_{P,c}$ with respect to entropy in Fig. \ref{RHCP1}.
 It can be observed that there exists a one-to-one correspondence between the singularities of the Ricci scalar $R^{H}$ and the phase transition of $C_{P,c}$.

Fig. \ref{RHCP1} also reveals that, for $P\neq P_{c}$ there exists two divergent points for $C_{P,c}$. The unstable phases with negative specific heat happen in the lower and higher entropy regions, while the intermediate phase with positive $C_{P,c}$ value is stable phase. As temperature is lower or higher than its critical value there are three possible phases, i.e., the cold black hole (CBH), intermediate black hole (IBH) and hot black hole (HBH). As temperature reaches to its critical value $T=T_{c}$, these two divergent points get closer and coincide at $\hat{S}=1$ to form a single divergence where the stable region disappears. It means that the black hole is unstable for $P=P_{c}$. 

\begin{figure}[t]
\includegraphics[trim={1mm 1mm 0 0},clip,scale=0.53]{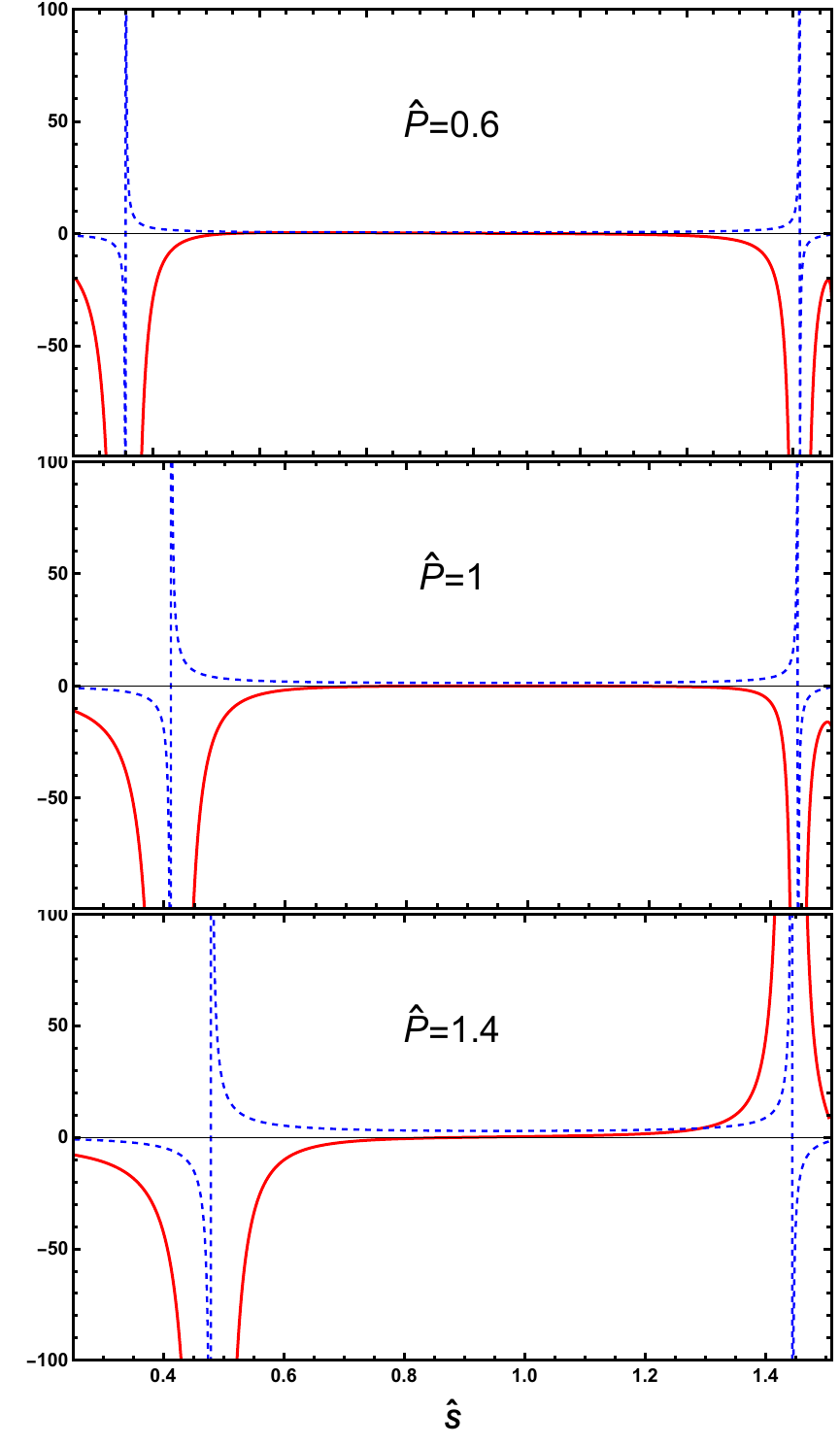}
\caption{The diagram of the specific heat $C_{V,c}$ (blue dashed curve) and  curvature $R^{G}$ (red curve) versus the entropy $\hat{S}$ for $\hat{P}=0.6$, $\hat{P}=1$, and $\hat{P}=1.4$. Note that the same figures can be obtined when one consider $R^{E}$. \label{RGCV1}}
\end{figure}
Before proceeding to the next section, allow us to examine the correspondence between the phase transitions of $C_{V,c}$ and singularities of the NTG curvature. To construct the appropriate NTG metric elements, we need to select either $\Xi= E(S,V)=M-PV$ or $\bar{\Xi}=G(T,P)=E-TS+PV=H-TS$ in Eq. \eqref{Ru1}, with the coordinates $(S,V)$ and $(T,P)$, respectively. In Fig. \ref{RGCV1}, $C_{V,c}$ and $R^{G}$ have been depicted with respect to entropy. It is obvious that there exists a direct mapping between the phase transition points of $C_{V,c}$ and the singularities of $R^{G}$.

A notable finding is that both heat capacities $C_{V,c}$ and $C_{P,c}$ are positive within the intermediate black hole (coexistence) region, as illustrated in Figs. \ref{RHCP1} and \ref{RGCV1}. This observation reinforces the argument put forth in \cite{Johnson:2019mdp} regarding the significance of considering the signs of $C_{P,c}$ and $C_{V,c}$ in assessing the stability of black holes within the extended framework.

\section{Coexistence curves and critical exponents}\label{sec3}

The aim of this section is to determine the critical exponents that describe the behavior of thermodynamic quantities such as heat capacities and thermodynamic curvatures, which were discussed in the previous section, near the critical point. In order to accomplish this, we construct the equal area law in the $\hat{T}-\hat{S}$ plane and then attempt to obtain the coexistence curve between the cold and hot black hole phases.\footnote{Alternatively, the phase transition can be obtained by analyzing the behavior of the Gibbs free energy, $G=H-TS$, as is usually done in studies of $PV$ criticality ---see, e.g., \cite{Kubiznak:2012wp}. In Fig. 2 of Ref. \cite{Johnson:2023dtf}, the Gibbs free energy of the qBTZ black hole is shown as a function of temperature. The swallowtail pattern, signifying a first-order phase transition, is observed for pressures both below and above the critical pressure. Thus, it is possible to determine the temperature and pressure of the phase transition, enabling the construction of the $\hat{P}-\hat{T}$ coexistence curve.}

The coexistence phase in $\hat{T}-\hat{S}$ plane occurs within a range of entropy values determined using the Maxwell equal area construction, namely\footnote{In the fixed charge ensemble the Gibbs free energy satisfies
\begin{equation}
dG=-SdT+VdP.
\end{equation}
Assuming that the ``Hot'' (H) and ``Cold'' (C) are two thermodynamic coexistence states of a first order phase transition, one easily has $\Delta G=G_{H}-G_{C}=0$. Integrating from state $C$ to sate $H$, we arrive at
\begin{equation}
-\int_{T_{C}}^{T_{H}}S dT+\int_{P_{C}}^{P_{H}} V dP=\int_{G_{C}}^{G_{H}}dG=0.
\end{equation}
For a constant $P$ curve in a $\hat{T}-\hat{S}$ diagram we can obtain the following equal area condition
\begin{equation}
\int_{T_{C}}^{T_{H}} S dT=0 \hspace{0.5cm} \text{or} \hspace{0.5cm} \int_{S_{C}}^{S_{H}} T dS=T_{\star} (S_{H}-S_{C}),
\end{equation}
 where we have integrated by parts and $T_{\star}=T(S_{C})=T(S_{H})$ is the temperature of the phase transition.}
\begin{equation}\label{maxwell}
 \hat{T}_{\star}\Big(\hat{S}_{H}-\hat{S}_{C}\Big)=\int_{\hat{S}_{C}}^{\hat{S}_{H}}\hat{T}d\hat{S},
\end{equation}
where $\hat{T}=\hat{T}_{\star}$ is the horizontal isothermal line. In this study, due to the complicated form of $T$, $S$, and $P$ (i.e., Eqs. \eqref{Tfunction}, \eqref{Sfunction}, and \eqref{Pfunction}, respectively), it is not possible to express temperature explicitly in terms of $(S, P)$. However, we can determine temperature as a function of entropy at a constant pressure using numerical methods.

\begin{figure}
\includegraphics[trim={1mm 1mm 0 0},clip,scale=0.6]{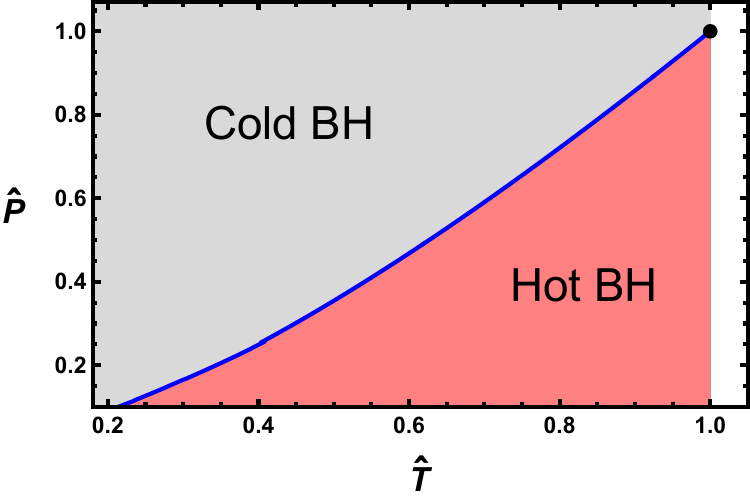}
\caption{Phases of the qBTZ black hole. The blue solid curve is the coexistence curve of cold and hot black holes. \label{PTdigram}}
\end{figure}

In Fig. \ref{PT1}, we have shown the location of $\hat{S}_{C}$ and $\hat{S}_{H}$ in the two black hole phases. Using Maxwell's construction \eqref{maxwell}, we have further plotted the $\hat{P}-\hat{T}$ phase diagram in Fig. \ref{PTdigram}. This diagram features the two distinct black hole phases: cold and hot black holes. The red solid curve represents the coexistence phase. This curve originates from the origin, gradually increases with temperature, and ultimately terminates at the critical point, which is denoted by a black dot.

Furthermore, the entropy difference is considered as an order parameter in $\hat{T}-\hat{S}$ plane, which quantifies the phase change across the critical point, using the Maxwell equal-area law.  Analogous to the Van der Waals fluid, we thus define
\begin{equation}
\eta=|\hat{S}_{H}-\hat{S}_{C}|.
\end{equation} 
Typically, near the critical point, the behavior of the entropy difference along the $\hat{P}-\hat{T}$ coexistence curve, is characterized by
\begin{equation}
\eta \sim \bigg\{ \begin{matrix}
(-t)^{\beta} & \hspace{0.5cm} \text{for} \hspace{0.5cm} t<0,\\
t^{\beta'} & \hspace{0.5cm} \text{for} \hspace{0.5cm} t>0,
\end{matrix}
\end{equation}
where $t=\hat{T}-1$.
Given the presence of a two-phase regime in the qBTZ system for either $t>0$ or $t<0$, it is reasonable to anticipate that the system exhibits identical critical behavior for both $t>0$ and $t<0$. 

\begin{figure}
\includegraphics[scale=0.55]{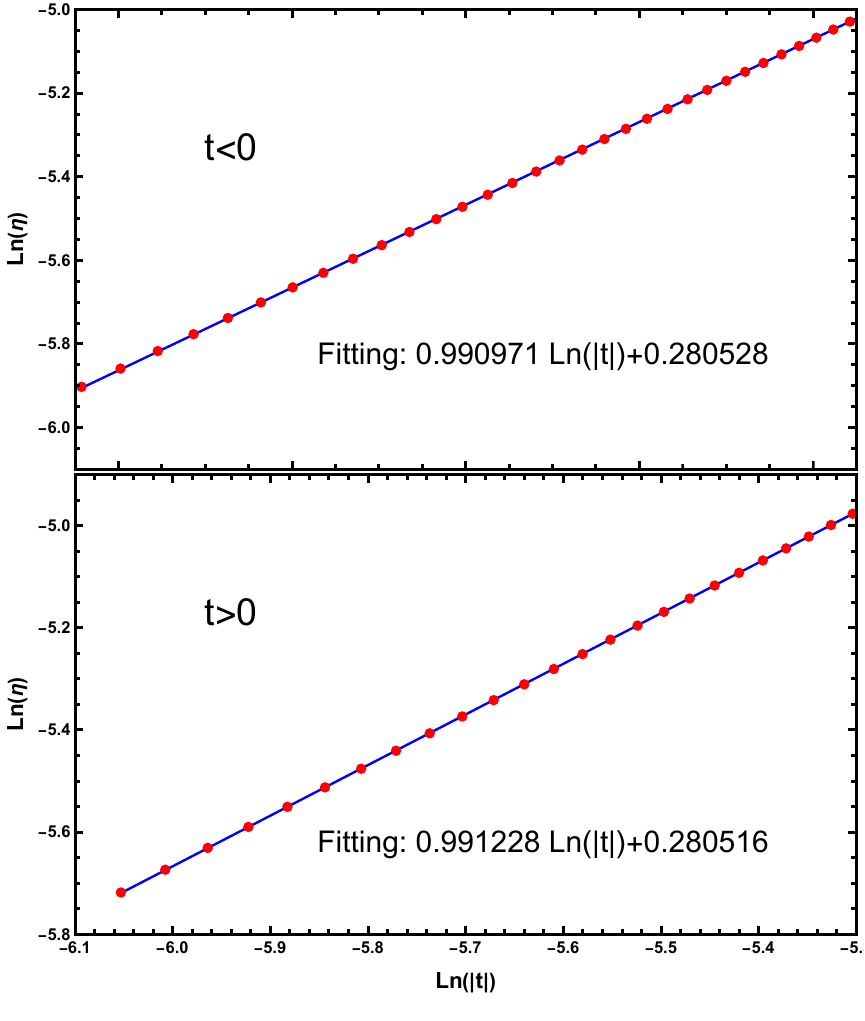}
\caption{Diagram of $\ln \eta$ versus $\ln|t|$ on crossing the $\hat{P}-\hat{T}$ cold-hot coexistence curve of the qBTZ black hole. The fitted straight line for the data points (red dot) is given by $\eta= b (-t)^{\beta}$ with $b=1.32383$ and $\beta=0.990971$ for $t<0$ and $\eta= b' t^{\beta'}$ with $b'=1.32381$ and $\beta'=0.991228$ for $t>0$. \label{etaTdigram}}
\end{figure}

\begin{figure}
\includegraphics[clip,scale=0.6]{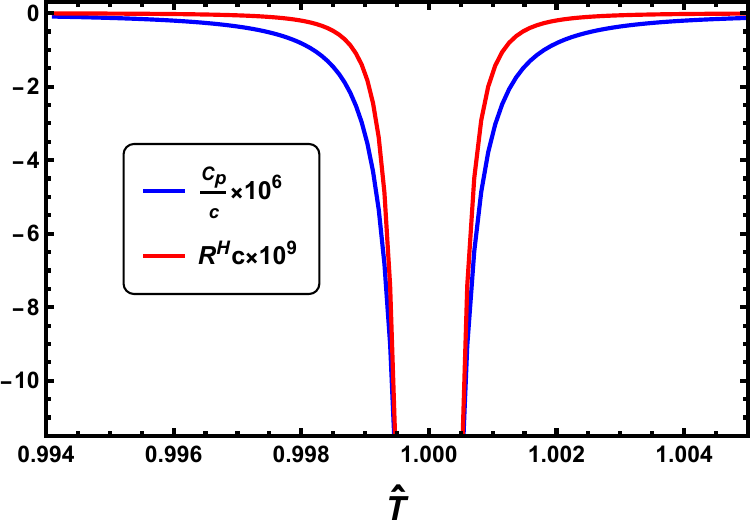}
\caption{The behaviour of $C_{P}$ and $R^{H}$ along the coexistence saturated cold curve. The same figure will be obtained for the coexistence saturated hot curve. \label{CPRHSL}}
\end{figure}

Fig. \ref{etaTdigram} presents the Ln-Ln plot of $\eta$ as a function of $t$ for $t\neq 0$. Form the slope and the intercept of the fitted straight lines for the numerical data points on the lower left with $|t|>10^{-4}$, we find
\begin{equation}
\eta \sim  |t| \hspace{0.5cm} \text{for} \hspace{0.5cm} t\neq 0.
\end{equation}
This shows that $\beta=\beta'=1$. These exponent values significantly differ from the classical mean-field critical exponents for the van der Waals fluid \cite{Niu:2011tb}.
In addition, the behavior of $C_{P,c}$ and $R^{H}$ along the coexistence curve and near the critical point are shown in Fig. \ref{CPRHSL}.  By increasing $\hat{T}$ for $\hat{T}<1$ or decreasing $\hat{T}$ for $\hat{T}>1$, both quantities decrease and diverge at the critical temperature  $\hat{T}=1$.
Near the critical point, one can consider the following critical behavior for $C_{P,c}$ and $R^{H}$ along $\hat{P}-\hat{T}$ coexistence curve.
\begin{equation}
\frac{C_{P,c}}{c}\sim \bigg\{ \begin{matrix}
(-t)^{-\alpha} & \hspace{0.5cm} \text{for} \hspace{0.5cm} t<0,\\
t^{-\alpha'} & \hspace{0.5cm} \text{for} \hspace{0.5cm} t>0,
\end{matrix}
\end{equation}
and
\begin{equation}
R^{H}c\sim \bigg\{ \begin{matrix}
(-t)^{-\gamma} & \hspace{0.5cm} \text{for} \hspace{0.5cm} t<0,\\
t^{-\gamma'} & \hspace{0.5cm} \text{for} \hspace{0.5cm} t>0.
\end{matrix}
\end{equation}
Employing Maxwell’s construction, we have studied the critical behaviour of $C_{P,c}$ and $R^{H}$ for $t>0$ and $t<0$ along $\hat{P}-\hat{T}$ coexistence curve, in Tab. \ref{tab1}.

\begin{table}[h]
\begin{center}
\begin{tabular}{c ||c||c||rr}
 \hline\hline
      Quantity & Coefficient   &  Temperature & SCBH & SHBH  \\\hline
 &  & for $t<0$& 2.013581 & 2.013420   \\
 & \raisebox{0.2cm}{$c_{C}$} &for $t>0$&  2.014214 & 2.012750  \\
\raisebox{0.2cm}{$\ln (|C_{P,c}|)$}
&  &for $t<0$& 1.794873 & 1.795891 \\
&\raisebox{0.2cm}{$d_{C}$}& for $t>0$& 1.787699 & 1.797094\\ \hline
& & for $t<0$  &  3.013427 & 3.013230  \\
& \raisebox{0.3cm}{$c_{R}$}& for $t>0$& 3.010046 & 3.010006 \\
\raisebox{0.2cm}{$\ln (|R^{H}|)$}
&  & for $t<0$& -0.285991 & -0.284792  \\ 
& \raisebox{0.3cm}{$d_{R}$}& for $t<0$ &-0.287398 & -0.287158 \\
\hline
\end{tabular}
\caption{Fitting values of the slope and intercept of $\ln |C_{P,c}|=-c_{C} \ln(|t|)+d_{C}$ and $\ln|R^{H}|=-c_{R}\ln(|t|)+d_{R}$ straight line for
  coexistence saturated cold black hole (SCBH) and coexistence saturated hot black hole (SHBH) curves.}\label{tab1}
\end{center}
\end{table}

According to the data shown in Tab. \ref{tab1}, by taking numerical error into account, one finds that 
\begin{eqnarray}\label{criticaltb0}
\nonumber  \frac{C_{P,c}}{c}(-t)^2 &=&-\exp \frac{ (1.794873+1.795891 )}{2} \approx -6.021,\\
\nonumber  \frac{C_{P,c}}{c} t^2 &=&-\exp \frac{ (1.787699+1.797094)}{2} \approx -6.003,
\end{eqnarray}
for $t<0$ and $t>0$, respectively. This implies $\alpha=\alpha'=2$ which are are not mean field critical exponents \cite{Niu:2011tb}. Similarly, one obtains
 \begin{eqnarray}\label{criticaltb0}
\nonumber  R^{H}c(-t)^3 &=&-\exp \frac{ (-0.285991-0.284792 )}{2} \approx -0.75171,\\
\nonumber  R^{H}c t^3 &=&-\exp \frac{ (-0.287398-0.287158)}{2} \approx -0.75030.
\end{eqnarray}
The above relations reveal that  $\gamma=\gamma'=3$. As a result, we have
\begin{equation}\label{criticalityrelation}
 \frac{C_{P,c}}{c} |t|^2 \approx -6 \hspace{0.3cm} \text{and} \hspace{0.3cm} R^{H} c|t|^3 \approx -\frac{3}{4} \hspace{0.5cm} \text{for} \hspace{0.5cm} t \neq 0  .
\end{equation}
Interestingly, the critical amplitudes are dimensionless constants and independent of the central charge. 
In addition, thanks to $dl_{H}^2=-dl_{F}^2$, we can conclude 
  \begin{equation}\label{RF}
 R^{F} C^{\rm qBTZ}_{V} |t|^3 \approx-2 \hspace{0.5cm} \text{for} \hspace{0.5cm} t \neq 0.
 \end{equation}
 where the heat capacity takes the positive finite value, i.e., $C^{\rm qBTZ}_{V} \approx 64 c \pi/75 $ near the critical point.
Furthermore, Eq. \eqref{RF} indicates a continuity in the criticality of thermodynamic curvature as we approach to the critical point for $t \to 0^{\pm}$ limits. 

This finding is significantly differ from the result reported in \cite{HosseiniMansoori:2020jrx,Wei:2019uqg} for a VdW fluid. In the VdW fluid case, we observe two-phase regime (the liquid/gas phase) only for $t<0$. 
Therefore, for $t<0$ we get close to the critical point along coexistence line between the liquid phase and the gas phase, whereas for $t>0$ we approach to the critical
point along the isochore line $v=v_{c}$ ($v$ is the specific volume) in $P-v$ plane. Consequently, our choice of
various trajectories in $P-v$ diagram on approaching to the critical point leads to a
discontinuity in criticality of the NTG curvature,  that is,
\begin{equation}
R^{F} C^{\rm VdW}_{V} |t|^2 \approx \bigg\{\begin{matrix}
-\frac{1}{8} & \text{for} \hspace{0.5cm} t<0,\\
-\frac{1}{2} & \text{for} \hspace{0.5cm} t>0,\\
\end{matrix}
\end{equation} 
 where $C^{\rm VdW}_{V}=3/2$ when one takes $k_{B}=1$ \cite{HosseiniMansoori:2020jrx,Wei:2019uqg}.

\section{Conclusions}\label{sec4}

In this paper we have carried out a thorough exploration of the critical behavior and extended phase structure of the quantum BTZ black hole. This solution was discovered via braneworld holography, and embodies the semiclassical backreaction effects of quantum conformal matter onto the geometry. Consequently, it serves as a holographic dual to a finite-$N$ quantum field theory.

Focusing on the fixed central charge ensemble, we first revisited the derivation of the heat capacities $C_{P,c}$ and $C_{V,c}$ using the bracket method introduced in \cite{Mansoori:2014oia}. Subsequently, we illustrated the phase structure of the qBTZ black hole in the $T-S$ plane. Notably, our analysis revealed a two-phase regime at both low and high temperature limits. Intriguingly, except for a critical pressure, $P=P_{c}$, the heat capacity $C_{P,c}$ undergoes a sign change at two phase transition points, indicative of the system transitioning between stable and unstable branches. Remarkably, $C_{V,c}$ remains positive at these points, allowing the system to seamlessly transfer between these branches. 

Utilizing the NTG geometry, we then formulated a metric to accurately represent the one-to-one relationship between phase transitions of heat capacities and the singularities of scalar curvatures. Employing this framework, we examined the critical behavior of heat capacities and scalar curvatures near the critical point. Specifically, we numerically assessed the criticality of such curvature along the coexistence curve in the $\hat{T}-\hat{S}$ diagram as it approached the critical point. 

Our findings revealed that the critical exponent of $C_{P,c}$ and the scalar curvature $R^{H}$ are 2 and 3, respectively, diverging from the well-known critical mean-field values for Van der Waals fluids. Other gravitational systems exhibiting non-standard critical exponents include the so-called Lovelock black holes \cite{Frassino:2014pha,Frassino:2016vww}. It would be interesting to understand more broadly the circumstances and mechanisms wherein mean field theory predictions may be circumvented. Notably, our numerical data unveiled that critical amplitudes remain unaffected by the central charge in the qBTZ black hole. This prompts an intriguing inquiry into their potential universality, which could be explored further by investigating other quantum black holes, such as the charged quantum BTZ \cite{Feng:2024uia,chargedEmparan}, and the quantum Schwarzschild-de Sitter and Kerr-de Sitter solutions in three dimensions \cite{Emparan:2022ijy,Panella:2023lsi}. We defer such investigation for future examination.\\

\noindent \emph{Acknowledgements.}
We are grateful to Roberto Emparan, Antonia Frassino and Andrew Svesko for useful discussions and correspondence. JFP is supported by the `Atracci\'on de Talento' program grant 2020-T1/TIC-20495, the Spanish Research Agency through the grants CEX2020-001007-S and PID2021-123017NB-I00, funded by MCIN/AEI/10.13039/501100011033 and by ERDF A way of making Europe.

\appendix

\section{Partial derivatives and bracket notation}\label{AppA}
Let us introduce the bracket method for doing partial derivatives. This allows us to calculate heat capacities very easily. 
 Generally, if we consider $f$, $g$, and $h$ as explicit functions of $(a,b)$, the partial derivative can be expressed by Poisson bracket as follows \cite{Mansoori:2014oia}:
 \begin{equation}
 \Big(\frac{\partial f}{\partial g}\Big)_{h}=\frac{\{f,h\}_{a,b}}{\{g,h\}_{a,b}}
 \end{equation}
 where $\{.,.\}$  is Poisson bracket given by
 \begin{equation}
 \{f,g\}_{a,b}=\Big(\frac{\partial f}{\partial a}\Big)_{b}\Big(\frac{\partial g}{\partial b}\Big)_{a}-\Big(\frac{\partial f}{\partial b}\Big)_{a}\Big(\frac{\partial g}{\partial a}\Big)_{b}
 \end{equation}
 
Moreover, we can generalize the bracket method for functions depending on $(n+1)$ independent variables. In this case, the partial derivative can be expressed as \cite{Mansoori:2014oia}
\begin{equation}\label{Poisson}
\Big(\frac{\partial f}{\partial g}\Big)_{h_{1},h_{2},...,h_{n}}=\frac{\{f, h_{1},h_{2},...,h_{n}\}_{q_{1},q_{2},...,q_{n+1}}}{\{g, h_{1},h_{2},...,h_{n}\}_{q_{1},q_{2},...,q_{n+1}}}
\end{equation}
where all $f$, $g$, and $h_{n}$ ($n=1,2,3,...$) are functions of $q_{i}$, $i=1,2,...,n+1$ independent variables and $\{.,.,.\}$ denotes Nambu bracket that is defined as,
\begin{equation}
\{f,h_{1},...,h_{n}\}_{q_{1},...,q_{n+1}}=\sum_{ijk...l=1}^{n+1} \epsilon_{ijk...l}\frac{\partial f}{\partial q_{i}}\frac{\partial h_{1}}{\partial q_{j}}\frac{\partial h_{2}}{\partial q_{k}}...\frac{\partial h_{n}}{\partial q_{l}}
\end{equation}
where $\epsilon_{ijk..l}$ is the Levi-Civita symbol.

\bibliography{CQBTZ}

\end{document}